\documentclass[longauth]{aa}  

\newcommand{\viny}[1]{{#1}}

\newcommand{\revise}[1]{#1}

\usepackage{graphicx}
\usepackage[varg]{txfonts}
\usepackage[T1]{fontenc}
\usepackage{aas_macros}
\usepackage{natbib}
\usepackage{subeqnarray}
\usepackage{color}
\usepackage{courier}
\definecolor{linkcolor}{rgb}{0.6,0,0}
\definecolor{citecolor}{rgb}{0,0,0.75}
\definecolor{urlcolor}{rgb}{0.12,0.46,0.7}

\def\ie{\emph{i.e.}}
\def\eg{\emph{e.g.}}

\def\T{\emph{(Top)}}
\def\B{\emph{(Bottom)}}

\newlength{\thsize}
\newlength{\hhsize}
\newlength{\khsize}
\newlength{\qhsize}
\def\d{{\rm d}}

\def\Cabspar{C_{\rm 1,abs}}
\def\Cabsper{C_{\rm 2,abs}}
\def\apeak{a_{\rm peak}}

\def\deg{^{\circ}}

\def\DUSTEM{\texttt{DustEM}}
\def\TMATRIX{\texttt{T-MATRIX}}

\begin{document}

\title{Polarized emission by aligned grains in the Mie regime : application to protoplanetary disks observed by ALMA}


\author{V. Guillet\inst{\ref{IAS},\ref{LUPM}}
, J. M. Girart\inst{\ref{ICE},\ref{IEEC}}
, A. J. Maury\inst{\ref{CEA},\ref{CfA}}
, F. O. Alves\inst{\ref{MPE}}
}

\institute{
\label{IAS}Universit\'{e} Paris-Saclay, CNRS, Institut d'astrophysique spatiale, 91405, Orsay, France \and
\label{LUPM}Laboratoire Univers et Particules de Montpellier, Universit{\'e} de Montpellier, CNRS/IN2P3, CC 72, Place Eug{\`e}ne Bataillon, 34095 Montpellier Cedex 5, France \and
\label{ICE}Institut de Ci\`encies de l’Espai (ICE, CSIC), Can Magrans S/N, E-08193 Cerdanyola del Vall\`es, Catalonia \and
\label{IEEC}Institut d’Estudis Espacials de Catalunya (IEEC), E-08034 Barcelona, Catalonia \and
\label{CEA}Laboratoire AIM, CEA/DSM-CNRS-Universit\'{e} Paris Diderot, IRFU, Astrophysics department, 91191 Gif-sur-Yvette, France \and
\label{CfA}Harvard-Smithsonian Center for Astrophysics, Cambridge, MA 02138, USA \and 
\label{MPE} Max-Planck-Institut für extraterrestrische Physik, Giessenbachstr. 1, D-85748 Garching, Germany
}

\date{A\&A accepted 22 January 2020}

 
  \abstract
{The azimuthal polarization patterns  observed in some protoplanetary disks by ALMA at millimeter wavelength have raised doubts about their being produced by dust grains aligned with the magnetic field lines. These conclusions were based on the calculations of dust polarized emission in the Rayleigh regime, \ie\ for grain sizes much smaller than the wavelength. However, the grain size in such disks is estimated to be typically in the range $0.1-1\,$mm from independent observations.}
{We study the dust polarization properties of aligned grains in emission in the Mie regime, \ie\ when the mean grain size approches the wavelength.}
{Using the \texttt{T-MATRIX} and \texttt{DustEM} codes, we  compute the spectral dependence of the polarization fraction in emission for grains in perfect spinning alignment, for various grain size distributions of weakly-elongated oblate and prolate grains of astrosilicate composition, with a mean size ranging from 10$\,\mu$m to 1 mm. }
{In the submillimeter and millimeter wavelength range, the polarization by B-field aligned grains becomes negative for grains larger than $\sim 250\,\mu$m, meaning that the polarization vector becomes parallel to the B-field. The transition from the positive to the negative polarization occurs at a wavelength $\lambda \sim 1$\,mm.
The regime of negative polarization does not exist for grains smaller than $\sim 100\,\mu$m.}
{When using realistic grain size distributions for disks with grains up to the submillimeter sizes, the polarization direction of thermal emission by aligned grains is shown to be parallel to the direction of the magnetic field over a significant fraction of the wavelengths typically used to observe young protoplanetary disks. This property may explain the peculiar azimuthal orientation of the polarization vectors in some of the disks observed with ALMA and attest of the conserved ability of dust polarized emission to trace the magnetic field in disks.}

\authorrunning{V. Guillet et al.}

\titlerunning{Polarized emission in the Mie regime}

\maketitle

%



\section{Introduction}



ALMA polarization observations have allowed to detect and resolve the linearly polarized dust emission at core (few thousands of au) and disk (few tens of au) scales.  
Thus, ALMA has revealed intriguing polarization features in the star forming cores, such as \viny{magnetized} accretion filaments and significantly polarized outflow cavities \citep{Maury18, LeGouellec19, Hull19}.  \viny{In the frame of the modern theory of grain alignment \citep{LH07}, these observations imply the presence of large (several tens of microns) grains in protostellar cores \citep{V19, Galametz19}} At disks scales, self-scattering of large grains, $\sim$100~$\mu$m, appears to be the  dominant mechanism in a significant fractions of young stellar disks 
 \citep{Kataoka2015, Kataoka16, Yang16, Yang17, Girart18, hull18, Bacciotti18, Dent19, Harrison2019,Sadavoy19}. 
 However, there are a number of sources where self-scattering can not explain the observed polarization properties. ALMA polarization observations of Class I objects such as HH 111 VLA 1 and [BHB2007] 11 can be interpreted as a combination of poloidal magnetic field morphology (produced by infall gas motions) and a toroidal one (produced by disk rotation) \citep[e. g.,][]{Lee18,Alves18}. The polarization levels observed for the latter is a at least factor of $\sim 2$ larger than self-scattering predictions. Therefore, \viny{a significant effort should be made to revise models of disk polarization in an attempt to match observations and satisfactorily both the dust and magnetic field properties in disks.}


\viny{Models of dust polarized emission that were designed for the interstellar medium \citep{DF09,Guillet2018} cannot be used to interpret observations of dust in protoplanetary disks where the grain size is a thousand times larger, typically of the order of a millimeter \citep{T19}.}
In this article, we model the emission of polarized radiation by aligned grains in the Mie regime, \ie\ when the grain size is of the order of the wavelength, and compare our predictions with polarization observations in protoplanetary disks.


This Letter is structured as follows: 
our dust model is presented in Section \ref{sec:model}. Section \ref{sec:obs} compares the predictions of our model with polarization observations in disks. Finally, Section \ref{sec:summary} provides a summary of our results.

\section{Polarized emission by aligned grains in the Mie regime}\label{sec:model}

In this section, we bear on the study by \cite{Guillet2018} to present a simple dust model allowing for the calculation of the polarized emission by aligned grains whose size approach the wavelength. 

\subsection{Dust model}

We model dust grains by oblate (disc-like) and prolate (needle-like) spheroids, spinning around their small axis which is their axis of maximal inertia. For simplicity, we assume that the orientation of their spin axis remains fixed in space (perfect alignment). 
We note $a$ the half-size of the symmetry axis, and $b$ the half-size of the two other axes.  
The axis ratio $b/a$ is larger than one for oblate grains, and smaller than one for prolate grains. 

The material composing the grain is another important ingredient in the calculation of dust emission. We use the so-called "astronomical silicates" \citep{DL84} characterized by a constant spectral index $\beta=2$ from the Far-Infrared to the millimeter domain. We restrict our analysis to homogeneous and compact grains. Calculations for composite fluffy aggregates with carbon/ferromagnetic inclusions and ice coating, while more relevant to disks, are beyond the scope of this first analysis of Mie polarized emission.



We use the \DUSTEM\ tool \citep{Compiegne2011,Guillet2018} to predict the polarized emission of a size distribution of grains. \DUSTEM\ consistently computes the equilibrium temperature of each grain size in any radiation field, and provides the spectral dependence of dust extinction and emission, polarized and unpolarized.
The grain absorption and scattering cross-sections were computed with the \TMATRIX\ code, a numerical adaptation of the Extended Boundary Condition Method \citep{Waterman1971} for non-spherical particles in a fixed orientation \cite{Mishchenko2000}. When the grain approaches the Mie regime, the grain cross-sections vary rapidly and strongly with the wavelength because of interferences between scattered waves \citep{Krugel}. To obtain a smooth spectral dependence of the grain optical polarization properties, calculations were done for 400 grain sizes between $0.3\,\mu$m and 3 mm. 

While the total intensity and polarized intensity strongly depends on the intensity of the radiation field heating the grains, its effect is much smaller on the polarization fraction in the submillimeter and millimeter wavelength range. Therefore, as a first step, we use the Interstellar Standard Radiation Field (ISRF) scaled with \viny{$G_0 = 100$, for which the grain temperature computed by \DUSTEM\ ($\sim 30\,$K for a $10\,\mu$m grain and $\sim10\,$K for a 1 mm grain) is within the range of current observational constraints \citep[\eg][]{T19}}.

\subsection{Effect of the size distribution with a magnetic field in the plane of the sky}\label{sec:sizedist}

We present the spectral dependence of the polarization fraction for different size distributions and inclination angle of the magnetic field.

\begin{figure}
\centering
\includegraphics[width=\hhsize]{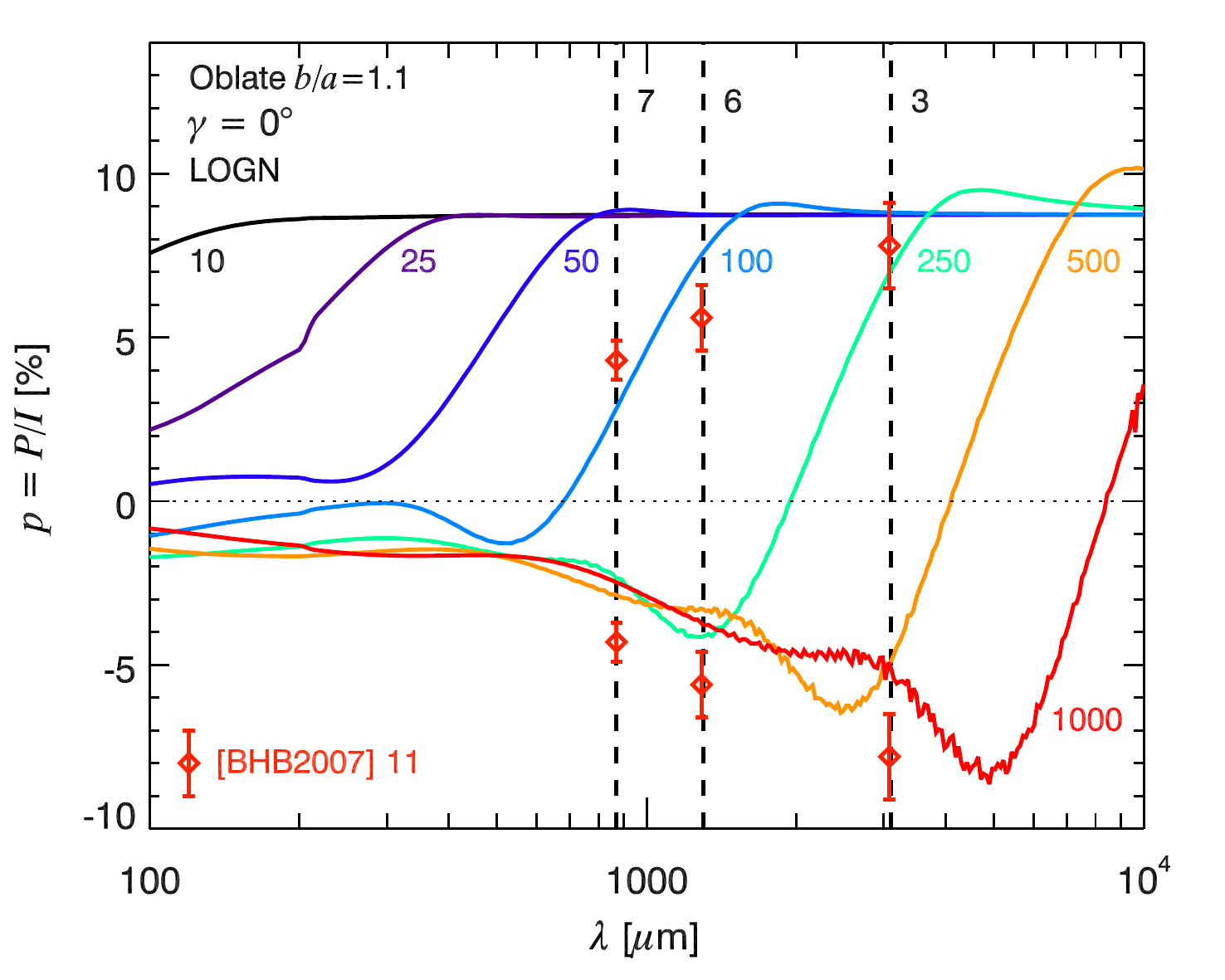}
\includegraphics[width=\hhsize]{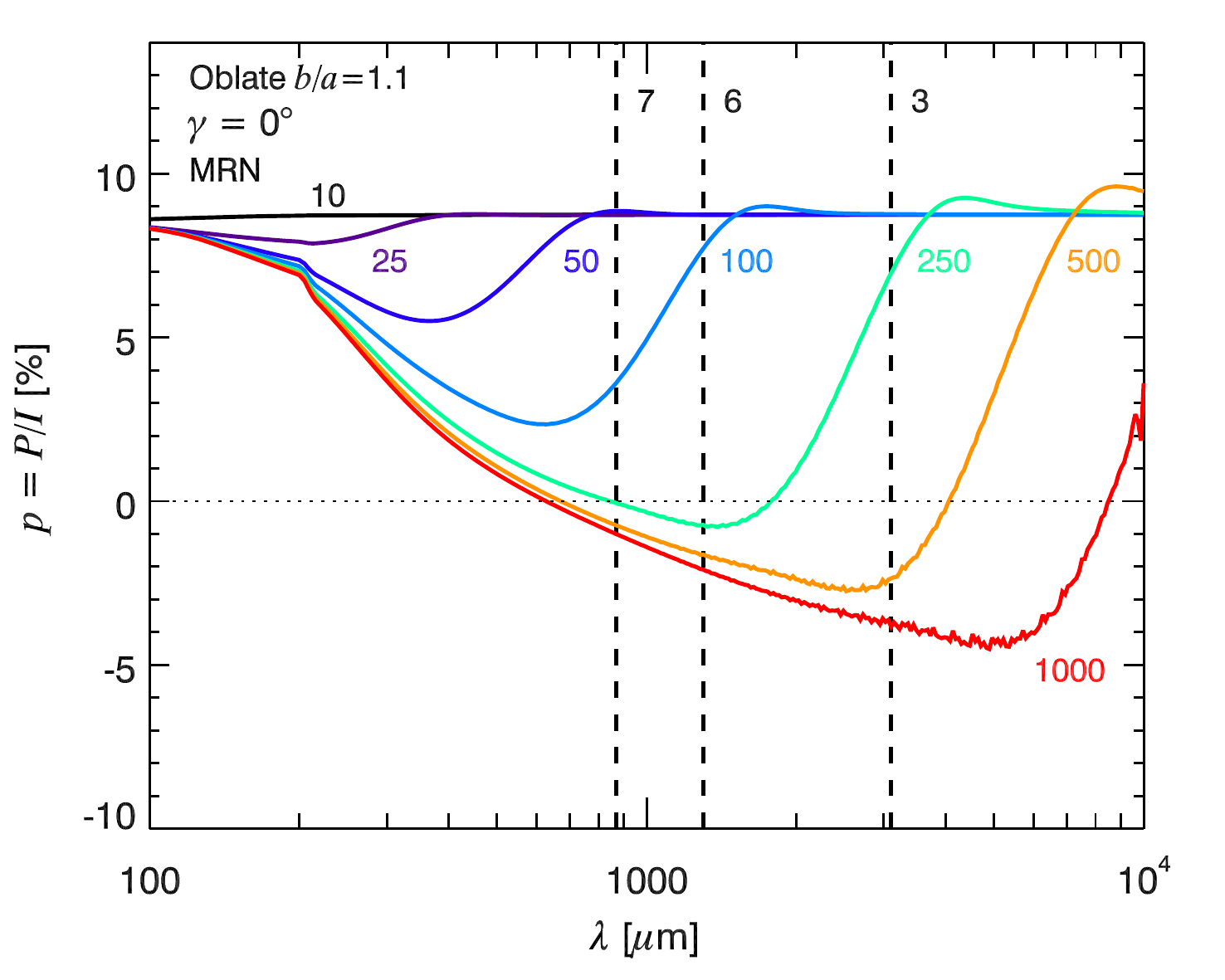}
\caption{
Polarization fraction in emission of oblate grains (axis ratio $b/a=1.1$), perfectly aligned in the plane of the sky. Results are presented for log-normal \T\ and MRN-like \B\ size distributions. The grain size corresponding to the peak of the mass distribution is indicated \revise{in $\mu$m}. Negative polarization fractions correspond to a rotation of the polarization vector by $90\deg$. The wavelength of ALMA bands 3, 6 and 7 is indicated with dashed vertical lines. Mean values of the polarization fraction observed for [BHB2007]~11 are overplotted in \viny{positive and negative} for the log-normal size distributions 
}
\label{fig:model}
\end{figure}


We first consider log-normal size distributions
\begin{equation}
\frac{\d n(a)}{\d a} \propto \frac{e^{-\log^2{(1.25\,a/\apeak)}/2\sigma^2}}{a}\,
\end{equation}
with $\sigma=0.1$, for different mean size $\apeak$ ranging from $10\,\mu$m to 1 mm. 
Such Narrow log-normal distributions are adapted to the modeling of single size distributions, and allows at the same time for the averaging of the physical oscillations of the grain cross-sections happening in the Mie regime (see Section \ref{sec:sizedist}).

The spectral dependence of the polarization fraction $P/I(\lambda)$ is presented in the top panel of Fig.~\ref{fig:model} for weakly elongated oblate grains ($b/a=1.1$). \viny{The polarization spectrum for prolate grains of comparable elongation ($b/a = 0.9$) is very similar and is, for the sake of brevity, hence not shown}.
For grains smaller than $10\,\mu$m, the polarization fraction $P/I$ is constant in the submillimeter and millimeter range, as expected for the Rayleigh regime ($2\pi a \ll \lambda$). 
As the grain size increases, this regime is shifted to longer and longer wavelengths. At shorter wavelengths, the polarization fraction is reduced and even negative for the larger grain sizes. A negative polarization fraction means that the direction of polarization of the grain thermal emission is parallel to the direction of alignment, and not perpendicular to it as in the Rayleigh regime. \viny{This means that weakly-elongated grains, rather counter-intuitively, emits on average preferentially with a polarization parallel to the minor axis than to the major axis. }. For $\apeak \ge 100\,\mu$m, the polarization fraction is negative from the submillimeter up to $\lambda_{\max} \simeq 2\pi\apeak$, peaking at a wavelength $\lambda_{\rm peak} \simeq 4\apeak$ before becoming positive again, increasing progressively towards the value characteristic for the Rayleigh regime. The larger the grains, the wider and deeper the regime of negative polarization.  
In Appendix \ref{sec:why}, we show that this specific property is actually not related to the grain size, but to the lower values of the imaginary part of the dust refractive index, $k(\lambda)$, in the submillimeter and millimeter wavelength range. The systematic decrease of $k(\lambda)$ with increasing $\lambda$ makes the transition from positive to negative polarization unavoidable. \revise{Note that this qualitative explanation is only valid for materials with a low value of the imaginary part $k$ of the refractive index, and can therefore not apply to the carbonaceous grains proposed by \cite{YL2019} to alleviate the tension between the grain size inferred from polarization and unpolarized observations in protoplanetary disks. Numerical calculations are needed to attest if the Mie polarized emission of such highly-absorbing materials is negative or not. This is part of a more general question of the impact of the grain shape, structure and composition on the Mie negative polarization regime, as we will emphasize in Sect.~\ref{sec:inclined}.} 

 
 
\viny{Power-law size distributions  may be more relevant than log-normal or single-size grain distributions to model what we can expect from a competition between coagulation and fragmentation \citep[\eg][]{Ormel2009}.} 
We now present our results for MRN-like \citep{MRN77} size distributions, completed by an exponential decay to insure a smooth upper cut of the size distribution. 
 \begin{equation}
 \frac{\d n(a)}{\d a} \propto a^{-3.5} \, e^{-\left(1.25\,a/a_{\rm peak}-1\right)^2}\,.
\end{equation}





The bottom panel of Fig.~\ref{fig:model} presents our results
for $\apeak$ ranging from $10\,\mu$m to 1 mm, for weakly elongated oblate grains with $b/a=1.1$. The polarization fraction in the submillimeter is now always positive whatever the size distribution, being dominated by the contribution of small grains which remain in the positive regime.
The negative regime, which appears only for $\apeak$ larger than a few hundred of microns, starts at a wavelength which does not vary much with $\apeak$, and is close to 1 mm. 
Further calculations (not shown here) indicate that the position of this transition wavelength also depends on the index of the power-law and on the value of the radiation field intensity, but only slightly. \viny{The spectral dependence of the polarization fraction for prolate grains with $b/a=0.9$ is very similar, and hence not shown.}

\subsection{Effect of grain shape when the magnetic field is inclined }\label{sec:inclined}

\begin{figure*}
\centering
\includegraphics[width=\hhsize]{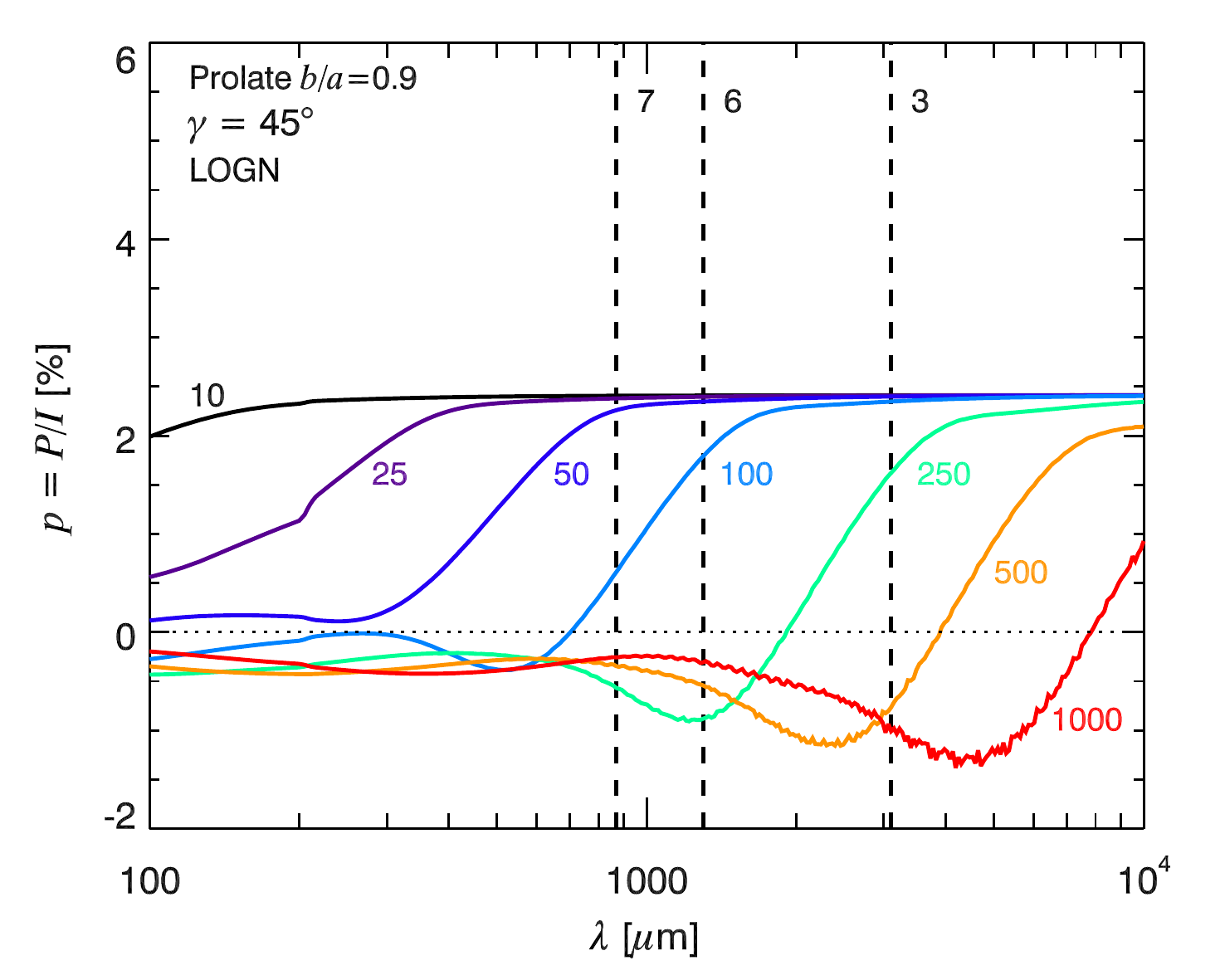}
\includegraphics[width=\hhsize]{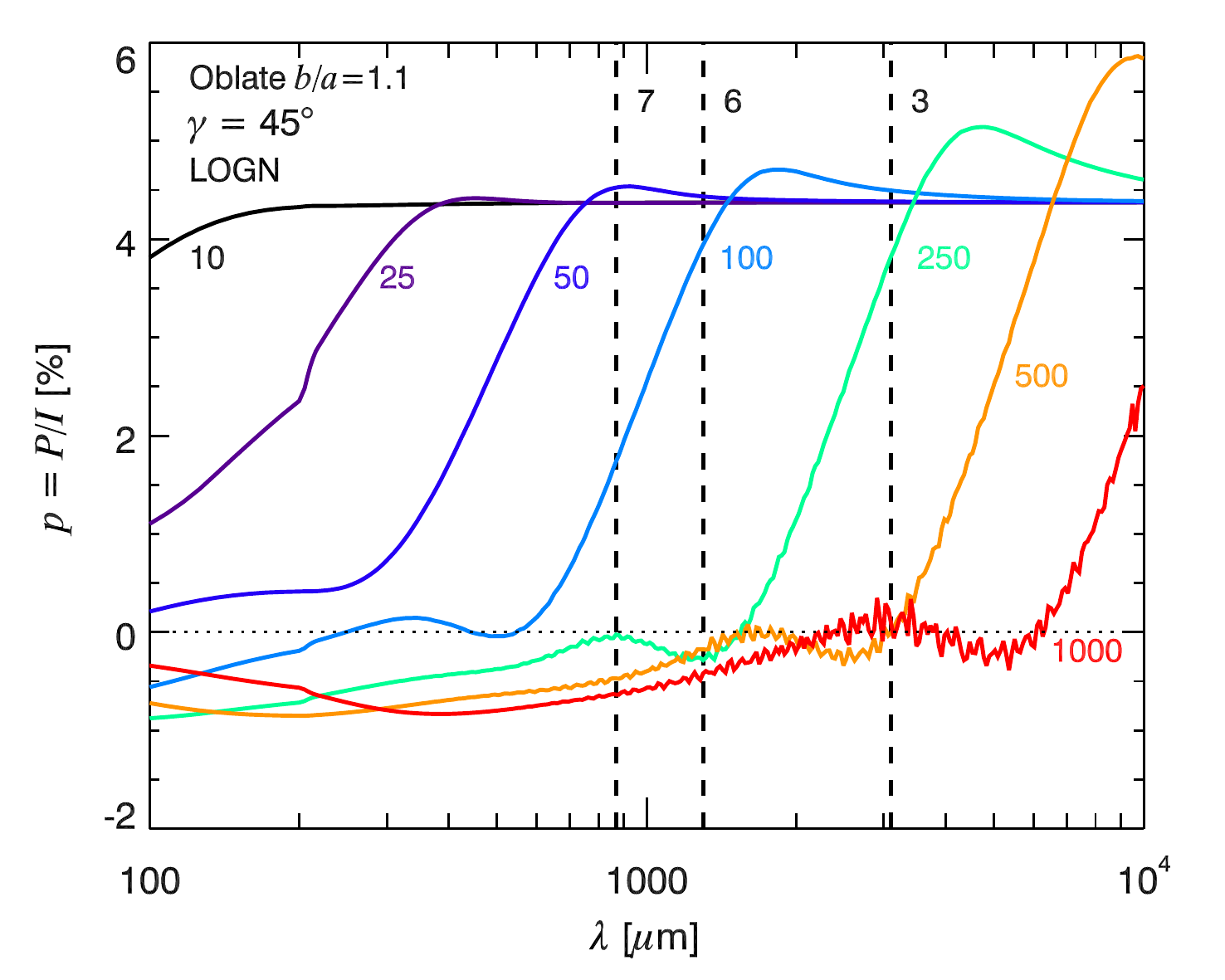}
\includegraphics[width=\hhsize]{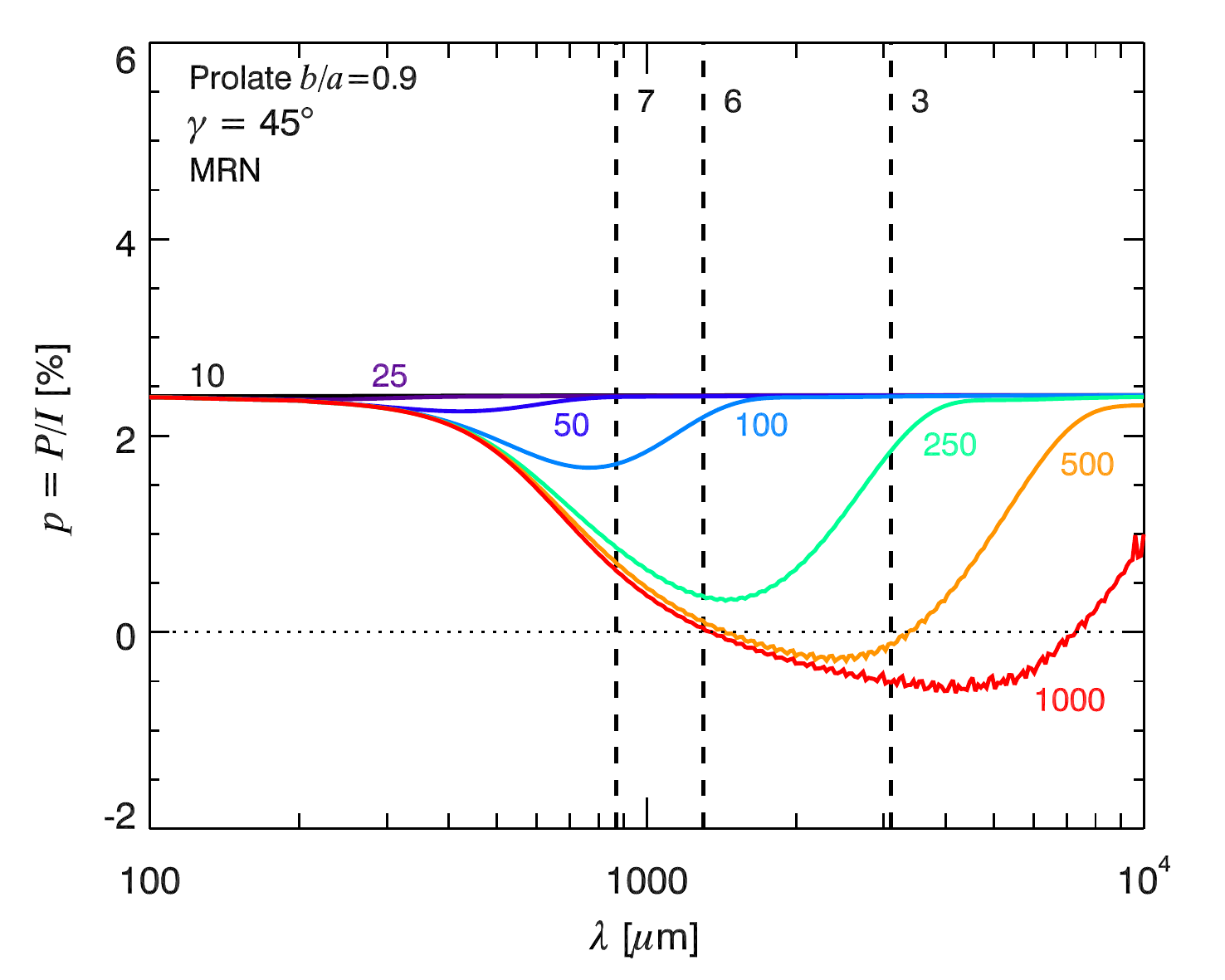}
\includegraphics[width=\hhsize]{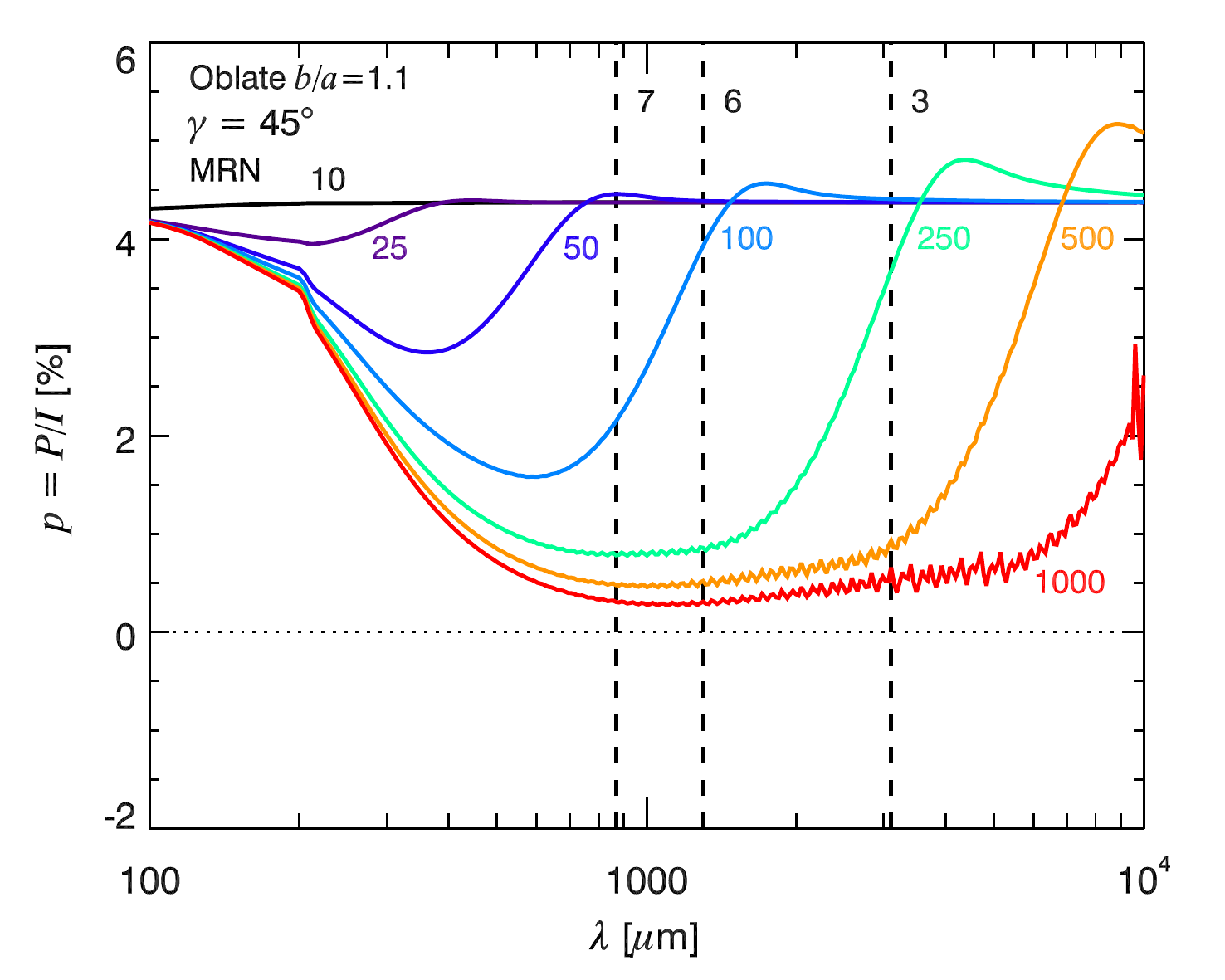}
\caption{\revise{Same as Fig.~\ref{fig:model}, for a direction of alignment inclined by $45\deg$ with respect to the plane of the sky. Log-normal \textit{(top row)} and MRN \textit{(bottom row)} size distributions of prolate \textit{(left column)} and oblate \textit{(right column)} grains.}}
\label{fig:model45}
\end{figure*}

\viny{The inclination of the magnetic field is a fundamental geometric ingredient in the modeling of polarized emission by aligned grains in the turbulent ISM \citep{Planck2018XII}, and  in protoplanetary disks \citep{Yang2019}. While the effect of the inclination of the magnetic field onto the polarization fraction is independent of the wavelength in the Rayleigh regime \citep{LD85}, it is not the case in the Mie regime \citep[\eg][]{Vosh89}, which may have some important consequences for the interpretation of polarization observations in inclined disks. 

} 



Fig.~\ref{fig:model45} shows the spectral dependence of the polarization fraction fo an inclination angle of $45\deg$ for weakly elongated prolate ($b/a=0.9$) and oblate ($b/a=1.1$) grains \revise{with log-normal and MRN size distributions}.
\revise{
For prolate grains with a log-normal size distribution, the $\gamma=45\deg$ case is similar to the $\gamma=0\deg$ case of Fig.~\ref{fig:model}, with a reduced amplitude. For oblate grains, the negative regime is more irregular and weaker. 
Due to the important weight of small particles in MRN size distributions, the negative polarization regime disappears in ALMA bands 7 and 6 in the prolate case, and does not exist at all in the oblate case.} 
This distinct behaviour of prolate and oblate grains, which did not appear when the magnetic field was in the plane of the sky (Sect. \ref{sec:sizedist}), is unusual in polarized emission. It points at the importance of the grain shape and of the grain spinning, precessing and nutating dynamics around the alignment direction in the Mie regime, and will therefore deserve more investigation.



\section{Applications to ALMA polarization observations of protoplanetary disks}\label{sec:obs}

In this Section, we compare our results to polarization observations of protoplanetary disks where the polarization patterns can not be explained by the self-scattering of the dust continuum.

\subsection{Azimuthal direction of polarization in protoplanetary disks}

The azimuthal direction of the polarization vectors observed in HL tau with ALMA at 3 mm, if interpreted as caused by aligned grains in the Rayleigh regime, leads to a configuration almost radial (poloidal) of the magnetic field which would be difficult to interpret \citep{Kataoka2017,Stephens2017}.
This anomalous polarization topology, which is not unique to HL Tau, raises doubts about the ability of dust polarization to trace the magnetic field in disks \citep{Stephens2017,Cox18,Sadavoy19}. Alignment by RATs along the radiation field anisotropy \citep{Tazaki2017} or by the gas flow \citep{Kataoka2019} have been invoked, although some of the detailed comparison to observations remain elusive \citep{Yang2019}. 


Negative polarization produced by $\gse 500~\mu$m grains may be a solution to the observed azimuthal polarization topology if we assume an azimuthal (toroidal) magnetic field, \revise{which is not unrealistic}. This specific property of Mie polarized emission has the advantage to be based on well-established optics and alignment physics with the magnetic field. \revise{Any departure from the expected azimuthal distribution would indicate a more complex magnetic field morphology. This is what may happen in the disk around the Class II AS 209 YSO \citep{Mori2019}, where the polarization distribution might not be related to the dust properties or alignment mechanism, but rather to the underlying structure of gas and dust rings to which the magnetic field is coupled.}
Our model also provides an explanation for the observed variations of the polarization fraction with the wavelength, for example in HL Tau between bands 3 and 7, while a pure Rayleigh polarization regime predicts a flat behaviour that is inconsistent with observations \citep{Yang2019}. \revise{If this scenario is correct, we expect to observe a flip of the polarization vectors by $90\deg$ at longer wavelengths (for $\lambda > 2\pi \apeak$) together with high polarization fractions ($\gse 10$\%), characteristic for the polarized emission by aligned grains in the Rayleigh regime (Fig.~\ref{fig:model}).}
Still, there remains to explain the lack of azimuthal variation in the polarization fraction at 3 mm that \cite{Yang2019} also pointed out. 
Negative polarization predicts a severe drop of the polarization \revise{fraction with $\gamma$}
(see Section \ref{sec:inclined}), which may not be compatible with ALMA observations. This will have to be investigated by a detailed model including the spectral dependence of polarization by self-scattering and polarized emission by aligned grains in the Mie regime.


\begin{figure}
\centering
\includegraphics[width=0.9\hhsize]{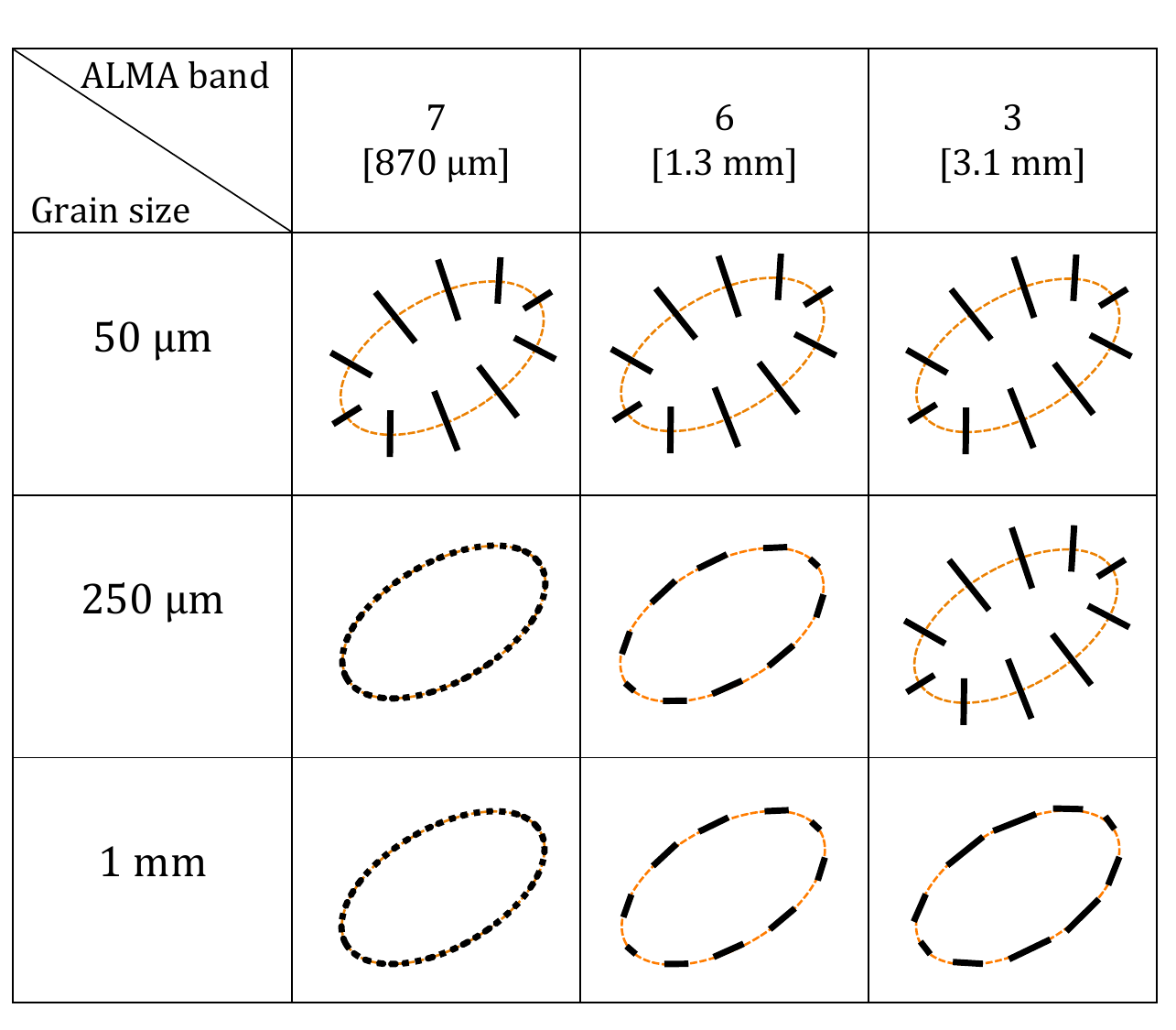}
\caption{\revise{Schematic representation of the polarization vectors from polarized emission by prolate ($b/a=0.9$) aligned grains that would be observed in the three ALMA bands for a protoplanetary disk inclined by an angle of $45\deg$ with a perfectly azimuthal magnetic field (dashed orange ellipses), for three values of the grain size corresponding to the peak of log-normal size distribution. The length of the vectors scales with the polarization fraction, reproducing the trends predicted by our model. Dotted lines indicate a polarization fraction that is close to zero. The contribution of self-scattering to the observed polarization is ignored.}}
\label{fig:schema}
\end{figure}

\revise{Fig.~\ref{fig:schema} schematically summarizes our results by showing the polarization vectors that, according to our model, should be observed in ALMA bands for grains of 50$\,\mu$m, 250$\,\mu$m and 1 mm, in the case of a protoplanetary disk inclined by an angle $45\deg$. 
For simplicity and clarity, this figure is based on the case of prolate grains with log-normal size distributions, which appears to be more regular. 
An azimuthal distribution of the magnetic field in the disk will be observed with an inclination angle ranging from $\gamma=0\deg$ along the minor axis to $\gamma=45\deg$ along the major axis of the disk.
In the Rayleigh regime ($a\ll \lambda$, top row), the direction of the polarization vector is perpendicular to the magnetic field, which is not exactly along the radial direction \citep[see the discussion in][]{Yang2019}, The polarization fraction is high, independent of the wavelength, and scales as $\cos^2{\gamma}$ \citep{DL84}. 
Contrarily, in the Mie regime of negative polarization (bottom row), polarization vectors are parallel to the magnetic field. The polarization fraction is lower than in the Rayleigh regime, decreases somewhat faster with $\gamma$ (Fig.~\ref{fig:model45}), and increases with increasing wavelength until it peaks (see Fig.~\ref{fig:model}). 
} 


%

Finally, we stress that the low levels of polarization from self-scattering of the dust grains \revise{(whose contribution is ignored in the schematic view of Fig.~\ref{fig:schema})}, that are widely-observed in protoplanetary disks around 1 mm, could be made possible because the otherwise much stronger polarization due to aligned grains cancels for large grains in a Mie regime around this wavelength (see Fig. 1 bottom).
\revise{The complication induced by superposition of the polarization signal by aligned grains and by self-scattering is underlined by the studies of \cite{Kataoka2016} and \cite{Ohashi2018} who analyzed the polarization patterns observed by ALMA band 7 at 870 $\mu$m in the protoplanetary disk around HD142527. To explain the change from a radial to an azimuthal polarization pattern from the southern to the northern part of the disk, these authors invoke an azimuthal magnetic field modulated by the presence of two different size distributions. Small ($\lse 100\,\mu$m) aligned grains in the Rayleigh regime would produce a high level ($\sim10\%$) of polarized emission in the South, while large ($\sim 100\,\mu$m) grains in the North would generate a low level of polarization (a few percents) by self-scattering. Similarly to the contradiction raised by \cite{Yang2019} for HL Tau, this scenario can not explain why the polarized emission by these large grains, much more intense than that produced by self-scattering, is not observed in the Northern part. This would impose  that such grains are not aligned, which is not compatible with the current view on grain alignment in disks \citep{HL16}. When the impact of the grain size on polarized emission is taken into accout, as is the purpose of this Letter, the contradiction however disappears. The large grains responsible for self-scattering contribute very little to the polarization observed at 870\,$\mu$m because it is around this wavelength that the transition from positive to negative polarization happens (Fig.~\ref{fig:model}). Even if the polarized emission by aligned grains is not strictly negligible in that particular region, its direction whether parallel (negative polarization) or perpendicular (positive polarization) to the azimuthal direction generated by self-scattering would not change the observed polarization pattern (azimuthal or radial), but would only result in a smaller or larger polarization fraction, respectively. 
}

\subsection{[BHB2007]~11}

The [BHB2007]~11 is protostellar binary system, classified as Class I \citep{Brooke17, Hara13}, with a relatively tenuous but elongated envelope surrounding a circumbinary disk, which is formed by a network of complex filaments \citep{Alves19}.  A molecular outflow appears to be launched at the centrifugal barrier, just outside of the circumbinary disk \citep{Alves17}. ALMA multi-wavelength polarization observations shows polarization within circumbinary disk \citep{Alves18}. Remarkably, the polarization directions are observed to be the same in bands 7, 6 and 3. The analysis of the polarization properties (polarization degree and pattern) indicates that this is produced by aligned grains with the magnetic fields instead of self-scattering and radiation fields. 

\viny{We built beam-matching maps obtained in the three bands using the same visibility range as in \citet{Alves18} but with a smaller robust weighting producing a common} synthesized beam of $0.20''\times0.15''$ with a position angle of $-80\deg$ for the three bands. We measured the polarization fraction in the three bands along the two main filaments detected by \citet{Alves19}, excluding the southern side, close to source B, with much larger values (but at the edge of the filament). The averaged polarization fraction is  
7.5$\pm$1.3\%, 5.7$\pm$1.1\% and 4.3$\pm$0.6\% at 3.1, 1.3 and 0.87~mm, respectively. \viny{Fig.~\ref{fig:model} compares these observations with our model predictions. In the absence of any flip of the polarization vectors between bands, the frequency dependence is incompatible with the regime of positive polarization, but is in rough agreement with the negative regime of large, almost single-sized, grains ($\gse500$~$\mu$m).} \viny{The presence of large grains in a Class I disk is not surprising as grains of several tens of microns seem to already be present in the youngest protostars \citep{V19}.}
The large values of polarization at the three frequencies suggests that the grain distribution should be closer to the log-normal rather than a modified MRN, and that grains should have more efficient polarization properties than the ones used here. A more detailed modeling in combination with observations at other wavelengths is needed to confirm this scenario, which would challenge the current interpretation of the magnetic morphology in this region \citep{Alves18}.

\section{Summary}\label{sec:summary}

\viny{The direction of the magnetic field is routinely assumed to be perpendicular to the polarization vectors of dust thermal emission, as expected in the Rayleigh regime of dust emission ($x  = 2\pi a \ll 1$). Our work show that this assumption may be incorrect when the size of the grain approaches the wavelength (Mie regime, $x \sim 1$). In the Mie regime, the polarization vectors becomes parallel to the magnetic field for weakly elongated spheroidal grains of prolate and oblate shapes, with size distributions either peaked or that follow a power-law.
This so-called negative polarized emission is present over a large range in wavelength that include ALMA bands. Polarized emission by aligned grains is predicted to become positive as soon as $\lambda > 2\pi a$. This can be tested with JVLA millimeter observations.
}


The polarization direction and wavelength dependence of the negative polarization in the Mie regime may reconcile some observations of protoplanetary disks such as HL Tau and [BHB2007]~11 with standard dust physics: dust grains are aligned with the magnetic field and their polarized emission still traces the magnetic field, though in an unsual way. The transition from the positive to the negative polarization regime, which for our dust model happens around $\lambda = 1$ mm for \revise{submm and} mm grains, may also explain why dust polarized emission by aligned grains is often not detected in disks.

Further studies are needed to assess how the detailed grain properties such as the shape, composition, fractal dimension and porosity will affect the properties of the negative polarization Mie regime, in particular its amplitude and its dependence with the magnetic field inclination.



\begin{acknowledgements}
We thank the anonymous referee for her/his constructive comments, suggesting Fig.~\ref{fig:schema} and enriching the discussion section. 
V. Guillet thanks S. Cabrit, A. Kataoka, H. Yang, F. M\'enard and G. Bertrang for stimulating discussions. 
J.M.G. is  supported by the Spanish grant AYA2017-84390-C2-R (AEI/FEDER, UE). 
F.O.A. acknowledges financial support from the Max Planck Society.
This work was funded by the European Research Council (ERC) under the European Union Horizon 2020 research and innovation programme (MagneticYSOs project grant agreement N.679937), and supported by the Programme National PCMI of CNRS/INSU.
This paper makes use of the following ALMA data: ADS/JAO.ALMA\#2013.1.000291.S and \#2016.1.01186.S. ALMA is a partnership of ESO (representing its member states), NSF (USA) and NINS (Japan), together with NRC (Canada) and NSC and ASIAA (Taiwan) and KASI (Republic of Korea), in cooperation with the Republic of Chile. The Joint ALMA Observatory is operated by ESO, AUI/NRAO and NAOJ.
\end{acknowledgements}

\bibliographystyle{aa}
\bibliography{biblio}

%
%
\appendix

\section{Why a negative polarization ?}\label{sec:why}

We have demonstrated that the polarized emission by aligned \viny{weakly-elongated} grains can present a regime of negative polarization when the grain size becomes of the order of the wavelength. In this regime, the direction of polarization is parallel to the alignment direction \viny{(\ie\ to the magnetic field)}, and not perpendicular to it as usually assumed. 

 Recently, \cite{Bertrang2019} independently found the signature of a negative polarization in the optical properties of 100$\,\mu$m prolate and oblate porous grains, composed of silicate, carbon or ice. 
These authors limited their calculations to grains smaller than $100\,\mu$m and did not include the calculation of the grain temperature. As a consequence, the negative polarization is absent from their prediction for the spectral dependence of the polarized fraction of an MRN-like size distribution which, consistently with Fig.~\ref{fig:model}, remained positive at all wavelengths. 

 \begin{figure}
 \centering
\includegraphics[width=\hhsize]{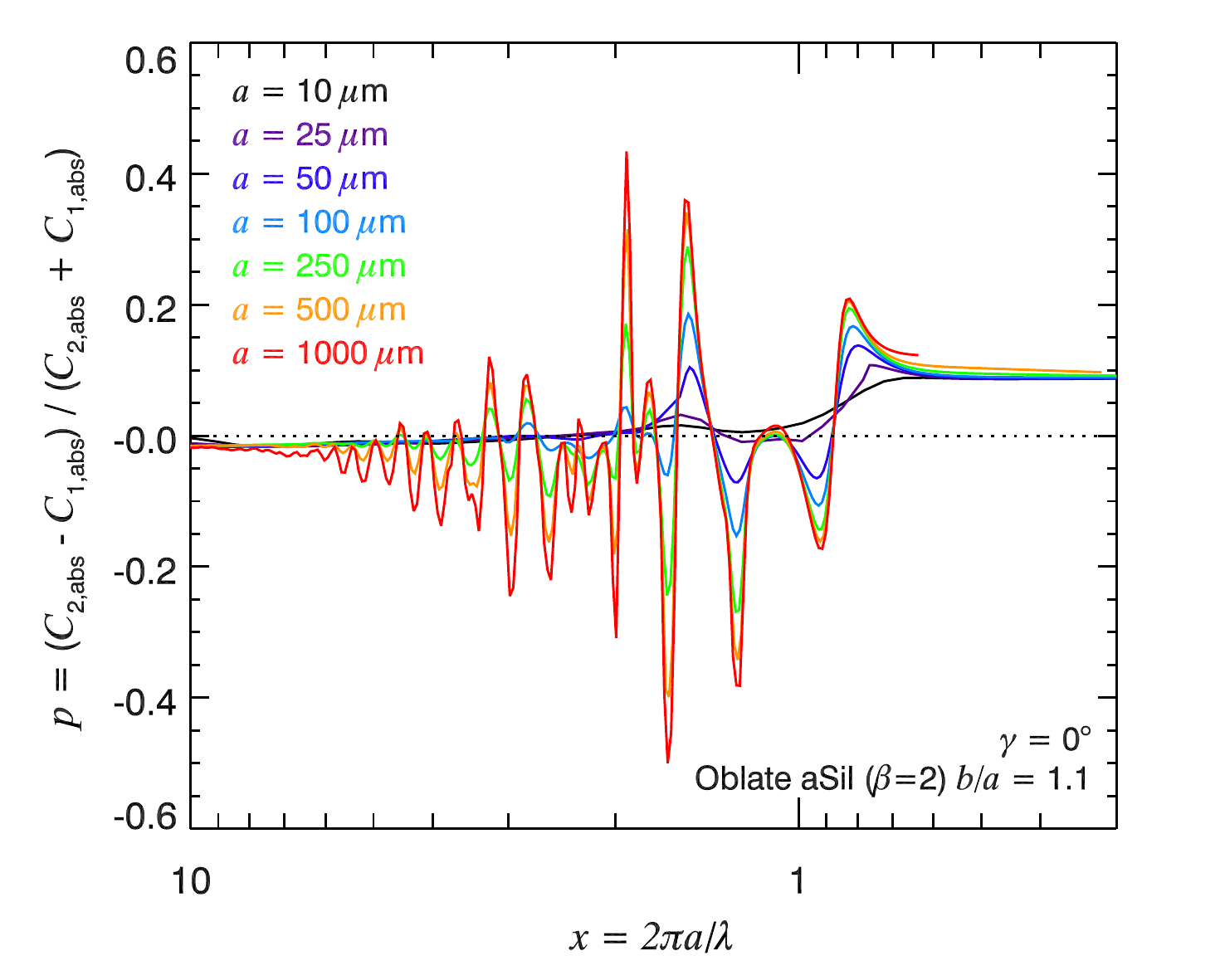}
\includegraphics[width=\hhsize]{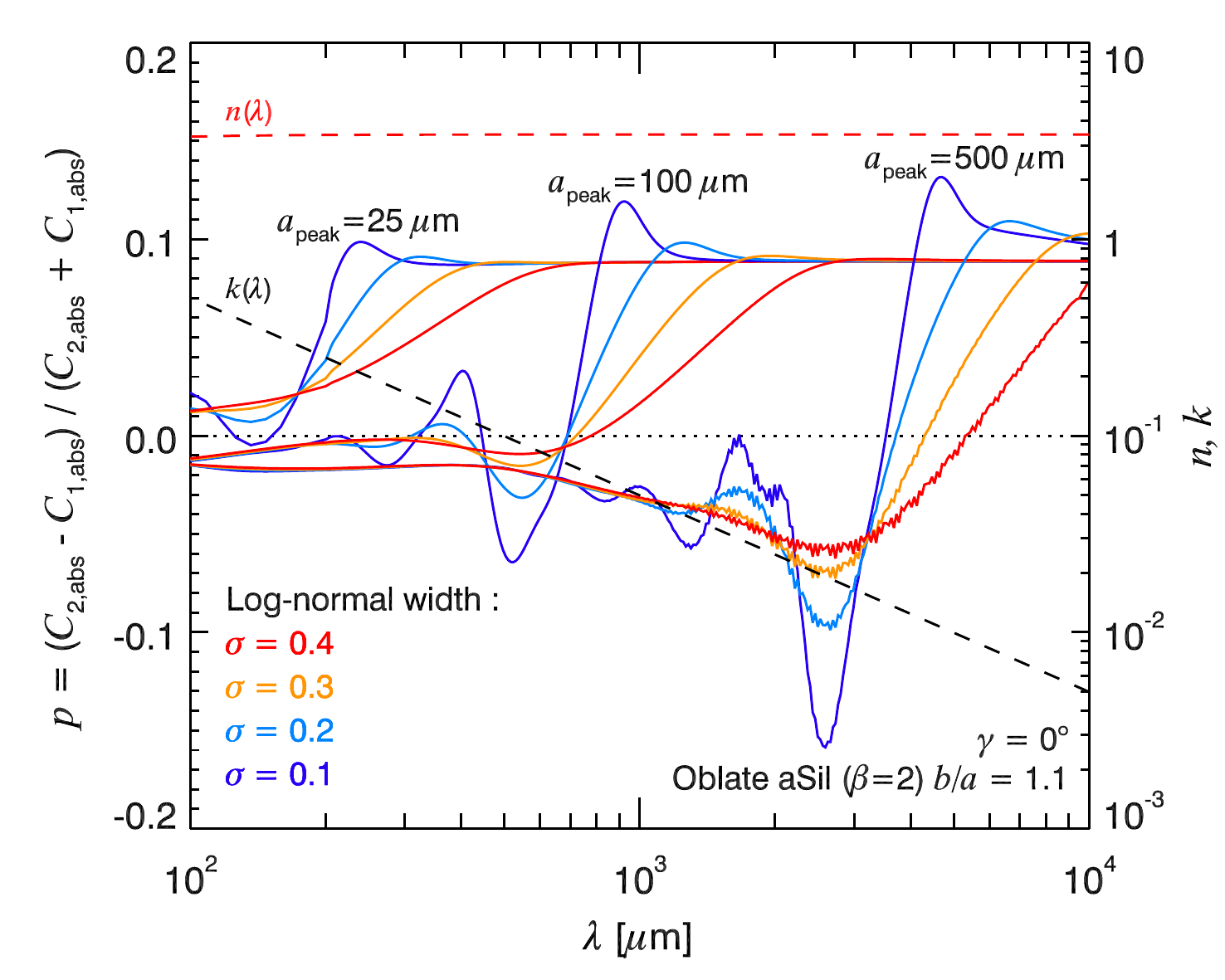}
\caption{\T\ Dust polarization efficiency $(\Cabsper-\Cabspar)/(\Cabsper+\Cabspar)$ as a function of the size parameter $x=2\pi a/\lambda$ for individual grains with a size comprised between 50 and $2000\,\mu$mm. Note that the $x$ axis has been reverted for an easier comparison with the bottom panel. \B\ Same as a function of the wavelength $\lambda$, for 3 mean grain size $\apeak$ 25, 100 and 500\,$\mu$m. The  grain of oblate shape has an axis ratio $b/a=1.1$ and is perfectly aligned. The direction of alignment lies in the plane of the sky ($\gamma=0\deg$).}
 \label{fig:PsI_lambda}
 \end{figure}
 
 According to Fig.~\ref{fig:model}, negative polarization does not happen for all grains sizes. To understand the reason for this, Fig.~\ref{fig:PsI_lambda} presents how the grain polarization efficiency vary with size size parameter $x=2\pi a/\lambda$.  \viny{The polarization efficiency of an individual , perfectly aligned, grain writes: $p=(\Cabsper-\Cabspar)/(\Cabsper+\Cabspar)$, where $\Cabspar$ (resp. $\Cabsper$) is the grain absorption/emission cross-section for an electromagnetic wave that is linearly polarized along (resp. perpendicular) to the direction of alignment projected onto the plane of the sky \citep[for more details, see][]{Guillet2018}.}
 We recall that for a given value of the complex refraction index $m=n+ik$, the dust optical cross-section only depends on the value of $x$, \ie\ on the ratio $a/\lambda$, and not on the value of $a$ and $\lambda$. \revise{Interferences between the waves scattered by the grain in the Mie regime produce strong oscillations \citep{Krugel}}. For $x \ll 1$, the polarization efficiency is positive, as expected in the Rayleigh regime. For $x \ge 1$, the polarization efficiency oscillates between positive and negative values, and converge toward zero at high $x$, as expected in the geometric optical regime where $a \gg \lambda$. We can observe two things. First, the value of $x$ for the zeros of the polarization fraction does not depend on the grain size. Second, the smaller the grain, the smaller the oscillations. As we comment below, this is explained by the variation of the complex refraction index with the wavelength.

The bottom panel of Fig.~\ref{fig:PsI_lambda} shows how the polarization fraction vary as a function of the wavelength, for different values of the width $\sigma$ of the log-normal distribution and mean grain $\apeak$. The real part $n$ and imaginary part $k$ of the optical constant for astronomical silicate are also plotted. The real part $n$, which controls the position of the zeros in the oscillations of the polarization efficiency \citep{Krugel}, is constant between $100\,\mu$m to 1 cm. The imaginary part, which controls the spectral dependence of the grain absorption cross-section and the amplitude of the oscillations in the Mie regime \citep{Krugel}, decreases as $k(\lambda)\propto \lambda^{1-\beta}\propto 1/\lambda$. The smaller grains ($25\,\mu$m here) are in the Mie regime in the far-infrared ($\lambda \sim 100\,\mu$m), where the absorption is still high ($k\sim 0.3)$. Larger values of $k$ tend to damp the oscillations (top panel of Fig.~\ref{fig:PsI_lambda}). The larger grains ($500\,\mu$m) fall in this regime in the millimeter, where $k$ has dropped by a factor 10, leaving stronger oscillations. These positive and negative oscillations average out when the size distribution gets wider (larger value of $\sigma$), with a positive mean value for smaller grains at small wavelength, and negative mean value for larger grains at large wavelengths. 


Our modeling was restricted to homogeneous, compact, weakly elongated grains. 
%
The detailed physical properties of dust aggregates (shape, composition, structure and fluffiness) are known to affect the absorption and scattering properties of dust grains. Their impact on  the regime of negative polarization deserves more investigation.

\end{document}